\documentclass[12pt]{iopart}

\usepackage{iopams}
\expandafter\let\csname equation*\endcsname\relax 
\expandafter\let\csname endequation*\endcsname\relax %
\usepackage{color}
\usepackage{graphicx}
\usepackage[utf8]{inputenc}
\usepackage[T1]{fontenc}
\usepackage{mathptmx}
\usepackage{enumerate}
\usepackage{theorem}
\usepackage{xcolor}
\usepackage{cite}
\usepackage{soul} 
\usepackage{physics}
\usepackage[justification=centering]{caption}

\newcommand{\spacesection}{\mathsf{S}} 
\newcommand{\metric}{\ensuremath{\mathrm{g}}} 
\newcommand{\metrics}{\ensuremath{\mathrm{h}}} 
\newcommand{\metricss}{\ensuremath{\mathfrak{g}}} 
\newcommand{\curvature}{\ensuremath{\mathsf{R}}} 
\newcommand{\gradD}{\raisebox{0.01em}{\scalebox{1.2}[1.0]{${\mathrm{D}}$}}}
\newcommand{\R}{\mathbb{R}}

\newcommand{\Hyp}{\mathbb{H}}
\newcommand{\Sph}{\mathbb{S}}
\newcommand{\normal}{\ensuremath{\mathrm{u}}}
\newcommand{\energyinner}{\raisebox{0.01em}{\scalebox{1.0}[1.4]{\ensuremath{\varepsilon}}}}
\newcommand{\eqstate}{\ensuremath{\mathrm{w}}}
\newcommand{\cell}{\ensuremath{\mathcal{K}}}
\newcommand{\hubble}{\ensuremath{\mathrm{H}}}
\newcommand{\scalef}{\scalebox{1.1}[1.1]{\ensuremath{\mathrm{a}}}}
\newcommand{\potential}{\scalebox{1}[0.8]{\ensuremath{\phi}}}
\newcommand{\tide}{\scalebox{0.8}[0.8]{\ensuremath{\mathfrak{M}}}}
\newcommand{\tidea}{\scalebox{0.8}[0.8]{\ensuremath{\widehat{\mathfrak{M}}}}}
\newcommand{\energy}{\scalebox{1.1}[1.1]{\ensuremath{\rho}}}
\newcommand{\pressure}{\scalebox{1.1}[1.1]{\ensuremath{\mathsf{P}}}}
\newcommand{\pnumber}{\scalebox{1.1}[1.1]{\ensuremath{\mathsf{n}}}}
\newcommand{\pnumberv}{\scalebox{1.1}[1.1]{\ensuremath{\hat{\mathsf{n}}}}}

\usepackage{soul} 
\soulregister\cite7 
\soulregister\ref7 
\soulregister\eqref7 

\begin{document}

\title[Decelerating models with large-scale acceleration]{A class of decelerating inhomogeneous cosmological models giving rise to accelerating FLRW universes at large scales}

\author{Leandro G. Gomes}
\address{Federal University of Itajub\'a (UNIFEI), Av. BPS, 1303, Itajub\'a-MG, 37500-903, Brazil}
\eads{\mailto{lggomes@unifei.edu.br}}

\begin{abstract}
In this manuscript, we develop a class of inhomogeneous relativistic cosmological models characterized by the following properties: (i) they admit cosmological observers for whom the spatial geometry and expansion are both homogeneous and isotropic; (ii) the matter content is an ensemble of massive particles with number conservation and responding viscously to the local tidal forces; (iii) on large scales, they reduce to FLRW cosmologies with dust and a dark‑energy term arising through backreaction; (iv) by appropriately choosing the current energy distribution, the large‑scale equation‑of‑state parameter reproduce any polinomial for an evolving dark-energy sector; (v) the luminosity distance derived from the large‑scale FLRW model implies acceleration—i.e., a negative deceleration parameter—even though the underlying inhomogeneous spacetime itself is decelerating. This apparent contradiction is resolved by noting that the global time coordinate employed in the large‑scale FLRW description cannot be identified with the proper time of any local observer experiencing homogeneous and isotropic expansion.
\end{abstract}


\maketitle

\section{Introduction}\label{sec:Intro}                                               
%
%

There have been quite many attempts to include the effects of local inhomogeneities on large-scale Cosmology beyond the perturbative treatment of the standard approach \cite{Book_2008_Weinberg}, with three distinguished fundamental aspects \cite{2011_RPP_Ellis_Clarkson}: averaging, backreaction, and fitting. General dynamical aspects of the averaged scalars \cite{Buchert1996A&A, Buchert2000GRG1, Buchert2001GRG2, Buchert2002GRG3}, the timescape approach \cite{Wiltshire2007NJP, Wiltshire2007PRL}, and the many different ideas of tilling space in congruent regions \cite{Wheeler_1957, Clifton_2012, Liu_2015, CliftonFerreira_2009, Zhuk_2015, bentivegna, Zhuk2021, Bruneton_2012,  Eingorn2021, Hellaby_2012, LGGomes_2022_CQG_2, LGGomes_2024_CQG_1} are among the main contributions in the field. We shall follow a middle ground between the standard cosmological and these inhomogeneous approaches. On one hand, we recognise the physical tenets underlying the former as the most powerful guide to cosmology; on the other, we introduce some ideas coming from the latter.

In this manuscript, we look for a class of models which keeps homogeneity and isotropy to the highest level, up to the point of allowing a non-trivial potential $\potential$ to appear, accounting for Newtonian-like gravitational interactions. We investigate the backreaction effects impinging on the FLRW models that the background spacetime naturally generates at large scales. In particular, we address the important issue of understanding how the observed accelerated expansion can be consistent with a decelerating universe. Although this seems to be contradictory at first glance, it is not at the fundamental level. We shall examine its origins and relations with similar ideas in the literature \cite{Wiltshire2007NJP, Wiltshire2007PRL}.     

The FLRW models are characterised by a set of cosmological observers, by which we mean the integral curves of a unit and vorticity-free time-like vector field $\normal$, the $\normal$-observers, who are free-falling, and whose spatial sections $\normal^{\bot}$ have homogeneous and isotropic geometry \cite{LGGomes_2024_CQG_1} experiencing an equally symmetric expansion \cite{LGGomes_2024_CQG_2}. In our model, we adopt all those assumptions but the ``free-falling" condition. This implies that anyone comoving with them fells a gravitational force $-\grad_{\normal}\normal$ in the $\normal$-observer perspective, and therefore, small volumes experience tidal forces as well. We set this framework in section \ref{Sec:Spacetime}. 

We focus on extending the dust FLRW solutions with the matter's reference frame decoupled from that of the $\normal$-observers, the FLRW being the sole case in which they meet. This means that the $\normal$-frame represents the locations where we can witness a homogeneous and isotropic universe, but not a frame where one can be at rest. Thus, matter is represented by an ensemble of moving particles of rest mass $m_0$, almost pressureless, except that they are still capable of resisting the gravitational tides by a viscosity term, which is not enough to halt their motion, but sufficient to keep the expansion homogeneous and isotropic. It is treated neither statistically nor as a fluid, but phenomenologically instead, with a conserved number density $4$-vector, $\pnumber^\mu$. 

In the context described so far, we analyse the evolution equations, namely, Einstein's with no cosmological constant ($\Lambda=0$) plus the conservation of particle number, and their solutions. This is the subject of section \ref{Sec:EvolutionEquations}. In the following, in section \ref{Sec:SolutionsPeriodicBoundary}, we introduce the notion of a periodic boundary condition, which is the mathematical apparatus ensuring that our model satisfies the cosmological principle on large scales. The existence and properties of the solutions therein are analysed, and the averaging problem is solved naturally. In section \ref{Sec:EffectiveLargeScaleModel}, we show that one of the backreaction effects is the rise of a term in the large-scale dynamics mimicking a positive and evolving dark-energy content, which is just an apparent situation, for it comes from the average value of the local gravitational potential. In section \ref{Sec:PhysicalDiscussion}, we discuss the physical implications and contextualization of our model. We address our final considerations in section \ref{Sec:FinalRemarks}.


\section{Spacetimes with a high degree of homogeneity and isotropy}
\label{Sec:Spacetime}

We begin by endowing our spacetime with a high degree of intrinsic symmetry \cite{coll_79, coll1_79,coll2_79}, so that its spatial geometry is homogeneous and isotropic \cite{LGGomes_2024_CQG_1}, as well as the expansion \cite{LGGomes_2024_CQG_2}. In the globally hyperbolic case, which we assume without loss of generality, it is the product manifold $I\times \spacesection$ \cite{CMP_2005_GlobalHyperbolic}, $\spacesection$ the simply connected space with Riemannian (time-independent) metric $\metricss$ of constant curvature $K$. Hence,  $\spacesection = \R^3$ is the Euclidean space, for $K=0$, $\spacesection = \Hyp^3$ is the hyperbolic space, for $K<0$, and $\spacesection = \Sph^3$ is the $3$-sphere, for $K>0$, as in the FLRW case. The cosmological observers ($\normal$-observers) are represented by the integral lines of the unit time-like vector field $\normal = e^{-\potential}\, \partial_t$ and the spacetime metric by
\begin{equation}\label{Eq:MetricConstantCurvature}
\metric = -\, e^{2 \potential} dt^2 + \scalef(t)^2\, \metricss \, ,
\end{equation}
where $\scalef(t)$ is the scale factor and $\potential: I\times \spacesection \to \R$ is the $\normal$-potential, uniquely defined, up to an additive time-function $\varphi(t)$ by the condition $\grad_{\normal}\normal_k=\partial_k\potential$ along the spatial directions \cite{LGGomes_2024_CQG_1}. Here, we use $\grad$ and $\gradD$ for the Levi-Civita connections of $\metric$ and $\metricss$, respectively. 

The $\normal$-observer at $x_0 \in \spacesection$ is represented by the time-like curve $(\tau, x_0)$, $\tau$ its proper time, such that $d\tau = e^{\potential(t,x_0)}dt$. Hence, except for the FLRW case, where we can set $\potential=0$, there is no natural choice for a global time function. This implies that, even though the expansion is homogeneous, the Hubble parameter is not, for it depends on the $\normal$-observer's proper time \cite{LGGomes_2024_CQG_2}, that is, 
\begin{equation}\label{Eq:HubbleParameter}
\hubble = \frac{1}{\scalef}\frac{d\scalef}{d\tau} = e^{-\potential}\, \frac{1}{\scalef}\frac{d\scalef}{dt} \, .
\end{equation}
Fortunately, we have the scale factor $\scalef(t)$ as a natural choice for the time coordinate,  which we use deliberately. In fact, the determination of the time parameter $t$ comes with the gauge choice for the potential $\potential$. For instance, assuming $\potential(\scalef,x_0)=0$ for the $\normal$-observer at $x_0 \in \spacesection$ is equivalent to setting $t$ as its proper time, so that  $\scalef(t)$ is determined through the relation \eqref{Eq:HubbleParameter} as far as $\hubble(\scalef,x)$ is given, that is, 
\begin{equation}\label{Eq:TimeDefinition}
t=t_0 + \int_{\scalef_0}^{\scalef(t)}\frac{d\scalef}{\scalef\, \hubble(\scalef,x_0)} \, .
\end{equation}
Notwithstanding, this is not the only way to set the time $t$. In section \ref{Sec:LargeScaleFRLW}, we give an explicit example of a gauge choice for $t$ that does not represent the proper time of any $\normal$-observer. Despite the nature of such a choice, we shall always set ``today" as $t=t_0$, with $t\to 0$ as $\scalef \to 0$, if applicable, and $\scalef(t_0) =1$. In this case, $t_0$ is the age of the universe according to the reference time $t$. Note that $\hubble$ and $\scalef$ are gauge-invariant, whereas $\potential$ and $t$ are not. Hence, we will often prefer the former over the latter.

\section{The matter content: The tidal-dust model}
\label{Sec:MatterContent}

There is no clear definition of what the interaction matter-gravity could be like in General Relativity. The best guess could be some non-minimal coupling in the Lagrangian, commonly discussed along with modified gravity \cite{capozziello2010beyond} and often followed by the breaking of some conservation law for matter, or some kind of interaction with gravitational energy and momentum in one of the forms of their quasi-local proposals, which usually depend on a coordinate choice or a specific boundary condition at spatial infinity (see \cite{QuasiLocalEnergyGravitation_LRR} and the references therein). Here, we develop a somewhat different phenomenological proposal based on the existence of the preferred $\normal$-frame, from where the expansion is homogeneous and isotropic, and using our Newtonian intuition to bring some classical ideas to our relativistic environment.

\subsection{General macroscopic aspects of an ensemble of self-interacting particles}
\label{Sec:ParticleConservation}

We assume matter to be a self-gravitating ensemble of particles whose energy density is composed of the ``rest" mass and some internal counterparts. For instance, we could realise one of these fundamental components as a typical galaxy of rest-mass $m_0$ and total energy
\begin{equation}\label{Eq:TotalEnergyParticle}
E=m_0 + E_{int}    \, ,
\end{equation}
the last term standing for the many contributions with kinematical, thermodynamical, or gravitational origin. The particle density at the spacetime point $(t,x) \in I \times \spacesection$ is $\pnumber(t,x)$, and the total energy density, as seen by the $\normal$-observers, is
\begin{equation}\label{Eq:EnergyDensity}
\energy 
= E\, \pnumber 
=  m_0\, \pnumber + \energyinner\, ,
\end{equation}
where $\pnumber(t,x)$ is the number density of particles and $\energyinner = E_{int}\pnumber$ is the density of internal energy. The number-flux $4$-vector is 
\begin{equation}
\pnumber^\mu = \pnumber\, \normal^\mu + \pnumberv^\mu \qquad (\pnumberv^\mu\normal_\mu =0)  \, ,    
\end{equation}
which is conserved, that is, 
\begin{equation}\label{Eq:ConservationParticleNumber}
\grad_\mu \pnumber^\mu=0 \, .
\end{equation} 
For this system, the part of the energy flowing through convection obeys the law
\begin{equation}\label{Eq:EnergyFlux}
q^i_{cv} =  E \, \pnumberv^i \, .
\end{equation}
The transport of energy is determined as we set the law for conduction underlying the system of particles, $q^i_{cd}$, so that the total flux of energy is $q^i= q^i_{cv}+ q^i_{cd}$.

As an example, if the ensemble of particles forms a fluid with unity $4$-velocity $\Tilde{\normal}_\nu$ and particle number density $\Tilde{\pnumber}$, as it is measured in its own reference frame, from the point of view of the $\normal$-observers, to whom the expansion is homogeneous and isotropic, its $4$-velocity is $\Tilde{\normal}_\mu = \gamma\, (\normal_\mu + v_\mu)$, $v_\mu$ the spatial peculiar velocity of the fluid, $\normal_\mu \, v^\mu =0$, and $\gamma = \left(1 - v^2\right)^{-1/2}$, the usual kinematical relativistic factor. Following it, in the $\normal$-frame, the number-flux $4$-vector is $\Tilde{\pnumber} \, \Tilde{\normal}^\mu = \pnumber \, \normal^\mu + \pnumberv^\mu$, so that
\begin{equation}\label{Eq:NumberFluxDensityDust}
\pnumber = \gamma\, \Tilde{\pnumber}
\quad \textrm{and} \quad  \pnumberv^i = \gamma\, \Tilde{\pnumber}\, v^i = \pnumber\, v^i \, .
\end{equation}
In the absence of viscosity and thermal effects, the energy-momentum tensor of a fluid is characteristic of dust, $T_{\mu\nu} = m_0\, \Tilde{\pnumber} \, \Tilde{\normal}_\mu \, \Tilde{\normal}_\nu$. In this case, the energy density, the energy-flux vector and stress tensor are, respectively, 
\begin{equation}\label{Eq:EMTDust}
\energy^{(dust)} = E\, \pnumber 
\quad , \quad
q_i^{(dust)} = E\, \pnumber\, v_i = E\, \pnumberv_i
\quad , \quad
T_{ij}^{(dust)} = E\, \pnumber\, v_i\, v_j \, ,
\end{equation}
with
\begin{equation}\label{Eq:TotalParticleEnergyDust}
E = m =\gamma\, m_0  = m_0\, \left( 1 + \frac{v^2}{2} + \ldots \right)\, .
\end{equation}
The internal energy of each particle composing the fluid is simply the relativistic kinetic energy of a typical particle, $E_{int}=m-m_0$.

\subsection{An analogy with Newtonian physics: gravitational forces}
\label{Sec:NewtonianForceTides}

Before we introduce our model for matter, we shall revisit some Newtonian ideas that lurk in our relativistic environment. First of all, whenever we choose a unit and vorticity-free timelike vector field $\normal$, as we did in section \ref{Sec:Spacetime}, the splitting of space from time following it brings us closer to the notion of an inertial frame in Newtonian mechanics: a reference frame from where the notions of spatial velocity and force arise. In particular, up to an additive time function, there is only one $\normal$-potential $\potential$ arising from the choice of the "frame" $\normal$ (see theorem 1 in \cite{LGGomes_2024_CQG_1}), which is the relativistic analogue of the Newtonian gravitational potential. Thus, seeing the choice of $\normal$ as that of an "inertial frame" and treating $\potential$ as the gravitational potential is, despite the clear limitations of this interpretation, a valuable and intuitive guide. 

Suppose a particle of rest mass $m_0$ is released by the $\normal$-observer at the instant $t=t_0$ at the point $x_0 \in \spacesection$. If no other forces act upon it, it is going to free-fall along a timelike geodesic, so that from the point of view of the $\normal$-observer at $x_0$, it instantaneously feels the gravitational force 
\begin{equation}\label{Eq:GravitationalForce}
F_i= -\, m_0\, \grad_{\normal}\normal_i = - \, m_0\, \gradD_i\potential 
\quad  \textrm{at $(t_0,x_0)$ } \, ,
\end{equation}
which resembles the Newtonian gravitational force. Following the same prescription, a volume element released in the same manner will face tidal forces due to the inhomogeneities of $\potential$. In fact, according to the geodesic deviation equation, with $s$ the affine parameter of the geodesic representing a free-falling particle released at $(t_0,x_0)$ (section 25.5 of Ref. \cite{Book_Thorne_ModernClassicalPhysics}), 
\begin{equation}\label{Eq:GeodesicDeviationEquation}
\frac{D^2\xi_k}{ds^2}= - \curvature_{k \mu \ell \nu}\normal^\mu  \xi^\ell \normal^\nu 
= -\, \tide_{k\ell}\, \xi^\ell 
+ \left(  e^{-\potential} \frac{\partial \hubble}{\partial t} + \hubble^2  \right)\, \xi_k \quad  \textrm{at $(t_0,x_0)$.}
\end{equation}
Here, $\tide_{\kappa\lambda}$ is the Newtonian tidal tensor of the $\normal$-observers, that is, the unique tensor field vanishing along time directions, $\tide_{0\mu}=\tide_{\mu 0}=0$, whether along the spatial ones, 
\begin{equation}
\tide_{ij}=\gradD_i\gradD_j\potential+ \gradD_i\potential\gradD_j\potential \, .
\end{equation}
This definition is justified by the fact that: (a) It depends only on $\normal$, for $\potential$ is defined up to an additive time function \cite{LGGomes_2024_CQG_1}; (b) It clearly reproduces the same concept of Newtonian physics in the limit $|\potential|<< 1$ \cite{Book_2013_Ohanian}; (c) Equation \eqref{Eq:GeodesicDeviationEquation} is exactly the relativistic version for tidal forces, so that $\tide_{k\ell}$ plays an analog role to its Newtonian version. Note that the second term in equation \eqref{Eq:GeodesicDeviationEquation} is purely relativistic, representing the effect of the cosmological expansion on the lengths and vanishes as $\hubble = 0$.

We can decompose the Newtonian tidal tensor into its bulk and anisotropic components, as
\begin{equation}
\tide_{\mu\nu} = \tide\, \left(\metric_{\mu\nu}+\normal_\mu\normal_\nu\right) + \tidea_{\mu\nu}\, ,
\end{equation}
where, using the notation $\gradD^2f$ for the Laplacean and $|v|=\sqrt{\metricss_{ij}v^iv^j}$ for the norm of the metric $\metricss = \metricss_{ij}dx^idx^j$, we have 
\begin{equation}
\tide =\frac{1}{3} \tide^\mu_\mu = \frac{1}{3\, \scalef^2}\, \left(\gradD^2\potential+ |\gradD\potential|^2\right) 
\quad \textrm{and} \quad  \tidea^\mu_\mu =0 \, .
\end{equation}
Note that in the Newtonian limit $|\potential|<<1$, the tide pressure $\tide$ is $1/6$ the mass density, a consequence of Poisson's equation, and therefore non-negative.

\subsection{A phenomenological approach to the tidal-dust model}
\label{Sec:TidalDustModel}

We shall deal with the description of matter content in the $\normal$-observers point of view with some kind of gravitational self-interaction taken into account. The fluid picture might not be suitable for backreaction in the homogeneous and isotropic expansion context, as it is indicated by the no-go theorem of Buchert-Ehlers \cite{Buchert1996A&A,Buchert:2017obp} and some numerical simulations \cite{macpherson2019einstein}, while kinetic theory is surrounded by technical issues that make it prohibitive for a first approach (see section 5.4 of Ref. \cite{Book_2012_ellis_mac_marteens} and references therein). Since our aim is to investigate the behaviour of inhomogeneities under homogeneous and isotropic expansion beyond the perturbative approach of the standard model, which has already been well established, we might look for an intermediate picture for matter, which is at the same time simple and physically relevant. Hence, we closely follow the phenomenological aspects of the theory of transport phenomena, where the coefficients, such as viscosity, are introduced by hand under suitable physical justifications. 

Our model for matter is an ensemble of classical particles interacting viscously with the Newtonian-like tidal forces they generate. Instead of the fluid picture, where the number-flux density determines the fluid's velocity as in \eqref{Eq:NumberFluxDensityDust}, here a swarm of particles executes random motions around their local centre of mass, ensuring that no relative velocity is privileged within this frame. We neglect any thermodynamic effects, such as heat conduction and thermal pressure, for the sake of simplicity only, leaving this more realistic and elaborate situation to be investigated elsewhere. Hence, following these prescriptions, we assume the total energy flux-density to be convective and ruled by the general law \eqref{Eq:EnergyFlux}, that is,    
\begin{equation}\label{Eq:EnergyFluxModel}
q_i =  E \, \pnumberv_i  \, .
\end{equation}

The determination of the stress tensor, as it is so long considered in elasticity with the elastic modulus tensor or in fluid dynamics with the bulk and shear viscosities \cite{Book_Thorne_ModernClassicalPhysics}, in cosmological models with imperfect fluids it is often assumed that matter responds linearly to the expansion, an effect usually interpreted as the viscosity in fluid dynamics, with bulk $-3\, \zeta\, \hubble$ and shear $- \lambda\,\sigma_{\mu\nu}$ components (Sec. 5.2 in Ref. \cite{Book_2012_ellis_mac_marteens}). In our context, we proceed with a similar approach guided by the notion of gravitational forces. We first note that the RHS of \eqref{Eq:GeodesicDeviationEquation} has a Newtonian-like term, $\tide_{ij}$, and an FLRW-like one, which, in the FLRW scenario ($\potential=0$), is proportional to $\ddot{\scalef}$. This cosmological term is not seen as a force in the Newtonian sense, but rather as the acceleration of spatial expansion. So, as we are interested in the gravitational interactions of the local universe, we are left with the former only.

In the local universe, at scales far smaller than the Hubble radius, the stress tensor of matter, $T_{ij}^{(L)}$, can be satisfactorily interpreted as its Newtonian analogue. Hence, if $T^{(LPF)}_{ij}$ denotes this stress tensor in the particle frame, then the tidal acceleration \eqref{Eq:GeodesicDeviationEquation} says that one reference frame is accelerated relative to the other, so that forces are different from these two points of view, and consequently the stress tensors must satisfy 
\begin{equation}\label{Eq:ChangeOfFrameStressTensor}
T_{ij}^{(L)} - T^{(LPF)}_{ij} \propto -\, \tide_{k\ell}  \, .
\end{equation}
We assume that the response of matter over the gravitational compression, stretching, and shearing obeys a linear law against the generating force, that is, $T^{(LPF)}_{ij}$ is linearly related to $\tide_{k\ell}$. Since no direction is favoured, the most general response is given by
\begin{equation}
{}^{(LPF)}T^{\ell}_{k} = \lambda_B\,  \tide \, \delta^\ell_k \, + \, \lambda_A\,  \tidea^\ell_k  \, ,
\end{equation}
where $\lambda_{B}$ and $\lambda_{B}$ are non-negative constants determining the viscosity effects. As we transcribe those relations to the $\normal$ frame using \eqref{Eq:ChangeOfFrameStressTensor}, the local stress tensor of matter is given by the relativistic pressure $\pressure = {}^{(L)}T^{\ell}_{\ell}/3$,
\begin{equation}\label{Eq:EMTPressure}
\pressure = -\, \zeta_{B}\,\tide  = -\, \frac{\zeta_{B}}{3\, \scalef^2}\, \left(\gradD^2\potential+ |\gradD\potential|^2\right)  \, ,
\end{equation}
where $\zeta_{B}$ is the tidal bulk viscosity, and the anisotropic stress $\pi_{\mu\nu}$, the traceless part of $T_{ij}^{(L)}$,
\begin{equation}\label{Eq:EMTAnisotropicStress}
\pi_{\mu\nu} =  -\, \zeta_{A}\, \tidea_{\mu\nu} \, ,
\end{equation}
with $\zeta_{A}$ the tidal anisotropic viscosity. Both coefficients are dimensionless and satisfy
\begin{equation}\label{Eq:ViscosityCoeficientsInequality}
0 \le \zeta_{A,B} \le \zeta_{max} \, ,
\end{equation}
where $\zeta_{max}$ is determined by the condition when the internal structure of matter offers no resistance against the gravitational forces acting on it, $\lambda_A=\lambda_B=0$, that is, 
\begin{equation}\label{Eq:ViscosityCoeficientsMaximum}
T^{(LPF)}_{ij} =0
\quad \iff \quad 
\zeta_{A} =  \zeta_{B} =  \zeta_{max} 
\quad \iff \quad 
{}^{(L)}T^{\ell}_{k} = -\, \zeta_{max} \, \tide^{\ell}_{k}\, .
\end{equation}
On the other hand, the condition
\begin{equation}\label{Eq:ViscosityCoeficientsRigidBody}
\zeta_{A} =  \zeta_{B} =  0 
\quad \Rightarrow \quad 
T^{(L)}_{ij} =0 \, ,
\end{equation}
says that the action of gravitation is cancelled out by the reaction of matter, so that the resultant forces acting upon matter, as seen by the $\normal$-observers, vanish. In section \ref{Sec:PhysicalDiscussion}, we return to the physical interpretation of these parameters.   

The final form of the total energy-momentum tensor is
\begin{equation}\label{Eq:EnergyMomentumTotal}
T_{\mu\nu} = \left(\, m_0\, \pnumber + \energyinner \, \right) \, \normal_\mu\normal_\nu 
+  E \, \left( \normal_\mu\pnumberv_\nu + \normal_\nu\pnumberv_\mu \right) 
-\, \zeta_{B}\,\tide\,  \left(\metric_{\mu\nu}+\normal_\mu\normal_\nu\right) 
-\, \zeta_{A}\, \tidea_{\mu\nu} \, .
\end{equation}
where $E$ is given in \eqref{Eq:TotalEnergyParticle} and the energy-flux density in \eqref{Eq:EnergyFluxModel}. Hereafter, we shall refer to this type of matter as tidal-dust, for it resembles the dust behaviour due to the absence of thermodynamical effects when matter is compressed by gravity, and it reduces to the dust FLRW model in the limit $\potential \to 0$. It represents a first step towards a more realistic model where thermal effects are taken into account.

\section{The evolution equations}
\label{Sec:EvolutionEquations}

%
%
\subsection{The complete set of equations}
\label{Sec:EinsteinEquations}

Our model has $6$ independent spacetime functions, namely, $\hubble$, $\energyinner$, $\pnumber^\mu$, and $2$ dimensionless parameters, $\zeta_A$ and $\zeta_B$. The behavior of our model is driven by Einstein's equations coupled to the energy-momentum tensor \eqref{Eq:EnergyMomentumTotal}, with no cosmological constant ($\Lambda=0$),  followed by the conservation of particle number \eqref{Eq:ConservationParticleNumber} and the relation \eqref{Eq:EnergyFluxModel}. They can be put in the following form \cite{LGGomes_2024_CQG_2}:  
\begin{enumerate}[(i)]
\item The generalised Friedmann equation is  
\begin{equation}\label{Eq:Friedmann}
\energy = 3\, \left( \hubble^2 + \frac{K}{\scalef^2}  \right)\, .
\end{equation}
\item Raychaudhuri's equation can be written as 
\begin{equation}\label{Eq:Raychaudhuri}
e^{-\potential}\, \frac{\partial\hubble}{\partial t} + \hubble^2  = 
- \, \frac{1}{6} \left( \energy + 3\, \pressure \right)
+ \frac{1}{3\, \scalef^2}\, \left(\, \gradD^2\potential + |\gradD\potential|^2\right)\, .    
\end{equation}
For our purposes here, we apply \eqref{Eq:EMTPressure}, \eqref{Eq:Friedmann}, and $e^{-\potential}\partial_t = \scalef\, \hubble\, \partial_{\scalef}$, to obtain
\begin{equation}\label{Eq:RaychaudhuriU}
\begin{split}
\frac{\partial U}{\partial s} = U^2 \left( D^2 U + \beta\, U \right)\, , 
\end{split} 
\end{equation}
which is derived just as in Ref. \cite{LGGomes_2025_PhySc_1},  with 
\begin{equation}\label{Eq:U}
U = \frac{1}{\scalef^{3/2}\, \hubble}
\quad , \quad s = \frac{1}{6}\, (2+\zeta_B)\, \left(1-\scalef \right)  
\quad \textrm{, and} \quad \beta = -\, \frac{3\,  K}{2+\zeta_B} \, . 
\end{equation}
\item The energy-flux equation with $q_k$ given in \eqref{Eq:EnergyFluxModel} implies
\begin{equation}\label{Eq:EnergyFluxEquation}
\pnumberv_k =-\, \frac{2\,\hubble}{E} \,  \gradD_k \potential 
=-\, \frac{2\,\hubble\, \pnumber}{\energy} \,  \gradD_k \potential \, .
\end{equation}
\item The traceless spatial equations:   
\begin{equation}\label{Eq:HubbleAnisotropyEquation}
\left( 1- \zeta_{A} \right)\, \tidea_{ij} =0 \, .
\end{equation}
It is a version of the tide-curvature-stress equilibrium \cite{LGGomes_2024_CQG_2,LGGomes_2025_PhySc_1}, with the anisotropic counterpart of the spatial Ricci tensor vanishing, since it comes from a space of constant curvature. 
\item The particle number conservation equation \eqref{Eq:ConservationParticleNumber} after substituting \eqref{Eq:EnergyFluxEquation}: 
\begin{equation}\label{Eq:ConservationParticleNumberU}
\frac{\partial \, \left(\scalef^3\, \pnumber\right)}{\partial \scalef} = 
2\, \gradD_k\left( \frac{\pnumber}{\energy}\, \gradD^k \potential\right) \, . 
\end{equation}
\end{enumerate}
\subsection{Remarks on the solutions of the evolution equations}
\label{Sec:RemarksGeneralSolution}
The first property of our equations is its FLRW limit as we set $\potential=0$. In this case, the metric \eqref{Eq:MetricConstantCurvature} and the energy-momentum tensor \eqref{Eq:EnergyMomentumTotal} become those of the dust FLRW spacetimes. For $\potential \ne 0$, we use the scale factor $\scalef$ as the time variable and define the smooth functions on $\spacesection$, $\pnumber_0(x)$ and $\energy_0(x)$, to be the initial conditions for the particle and energy densities today ($\scalef=1$), as 
\begin{equation}\label{Eq:InitialConditionParticle}
\pnumber(1,x) = \pnumber_0(x) \quad \textrm{and} \quad 
\energy(1,x)=\energy_0(x) \, .
\end{equation}
Assume that for any $x \in \spacesection$,
\begin{equation}\label{Eq:InitialConditionInequality}
\energy_0(x) > \max\{0,3\, K\} \, , 
\end{equation}
which is a very plausible assumption. In this case, we can use the auxiliary function $U$ and the simplifications it brings about. Its initial condition at $s=0$ ($\scalef=1$) is determined by the generalized Friedmann equation \eqref{Eq:Friedmann} as
\begin{equation}\label{Eq:InitialConditionU}
U(0,x) = U_0(x) = \sqrt{\frac{3}{\energy_0(x) - 3 K}} \, ,
\end{equation}
where the positive sign matches the expansion, that is, the initial Hubble parameter satisfies 
\begin{equation}
\hubble_0(x) = 1/U_0(x) >0
\end{equation} 
for all $x \in \spacesection$. Hence, with suitable boundary conditions, $U(s,x)$ can be determined from the Raychaudhuri equation \eqref{Eq:RaychaudhuriU}. Following it, we obtain $\hubble(\scalef,x)$ from \eqref{Eq:U}, and $\energy(\scalef,x)$ from \eqref{Eq:Friedmann}. The metric \eqref{Eq:MetricConstantCurvature} attains the form 
\begin{equation}\label{Eq:MetricConstantCurvatureU}
\metric = -\, \scalef\, U^2\, d\scalef^2 + \scalef^2\, \metricss_{ij}(x)\, dx^i dx^j \, ,
\end{equation}
while the gravitational potential $\potential$, following \eqref{Eq:HubbleParameter} and \eqref{Eq:U}, is
\begin{equation}\label{Eq:GravitationalPotentialU}
\potential = \ln\left( \scalef^{1/2}\, \dot{\scalef} \right) + \ln U\,\, ,
\end{equation}
which is fully determined only after the gauge choice for $t$. The particle density is determined by solving the equation \eqref{Eq:ConservationParticleNumberU} with the initial condition \eqref{Eq:InitialConditionParticle}. Note that the solution $U$ enters in \eqref{Eq:InitialConditionParticle} through $\energy$, except in the earlier epochs, when $\energy/\pnumber = E \approx m_0$ can be used, and $\pnumber$ can be readily integrated without $U$. 

The particle flux density is obtained from \eqref{Eq:EnergyFluxEquation}. Einstein's equations tell us that $\pnumberv$ represents the net flow of the local centre of mass falling under the local gravitational force according to \eqref{Eq:GravitationalForce}, that is, $\pnumberv \propto - \gradD\potential$. On the other hand, by assuming that the energy flows due to convection, according to our hypothesis \eqref{Eq:EnergyFlux}, we have also obtained 
\begin{equation}\label{Eq:EnergyFluxProportionality}
q_i \propto \pnumberv_i \propto - \gradD_i\potential \propto \gradD_i \hubble \, .
\end{equation}
The last relation follows from \eqref{Eq:HubbleParameter}, an interesting consequence of homogeneous and isotropic expansion, which demands energy to flow along the direction of maximum expansion.

The traceless spatial equations \eqref{Eq:HubbleAnisotropyEquation} are solved, in general, by setting the tide anisotropic viscosity as unit, that is, 
\begin{equation}\label{Eq:TideCurvatureStressEquilibriumZeta}
\zeta_A = 1\, .
\end{equation}
We address its physical meaning in section \ref{Sec:PhysicalDiscussion}.

Summing up the previous considerations, the solutions of the system of equations put forth in section \ref{Sec:EinsteinEquations} are obtained by setting the initial conditions \eqref{Eq:InitialConditionParticle} and the determination of $U$ from equation \eqref{Eq:RaychaudhuriU} with $\zeta_A = 1$ and $U_0(x)$ given in \eqref{Eq:InitialConditionU}. In the following, we shall introduce the asymptotic condition of a homogeneous and isotropic universe as $\scalef \to 0$. This allows us to infer interesting relations about the cosmological observable parameters.

\section{Solutions which are homogeneous and isotropic on large scales}
\label{Sec:SolutionsPeriodicBoundary}

\subsection{The periodic boundary conditions}
\label{Sec:BoundaryConditionPeriodic}

The concept of large-scale homogeneity and isotropy of the universe follows from the intuitive idea of tiling space into congruent parts of characteristic length $L_0$ and demanding periodic boundary conditions to Einstein's equations \cite{LGGomes_2022_CQG_2, LGGomes_2024_CQG_1, LGGomes_2025_PhySc_1}, so that it is homogeneous and isotropic as we look at it on scales much larger than $L_0$. We assume that all the geometric entities and their time derivatives are invariant by a free and properly discontinuous action on $\spacesection$ of a discrete group $\Gamma$ of $\metricss$-isometries whose quotient $\cell=\spacesection/\Gamma$ is a compact manifold of constant curvature $K$ and volume $L_0^3$ \cite{Wolf}. Here, $I \times \cell$ is called the cosmological cell and is naturally identified with the region in the spacetime $I \times \spacesection$ it represents. In this context, our spacetime satisfies a weaker version of the cosmological principle as compared to its usual implementation through the FLRW framework \cite{Book_1984_Wald}. A natural large-scale model emerges after taking spatial averages of the $\Gamma$-periodic scalars of the theory \cite{LGGomes_2024_CQG_1, LGGomes_2025_PhySc_1}, as
\begin{equation}\label{Eq:AverageValue}
\langle f \rangle (\scalef):= \frac{1}{L_0^{3}}\int_\cell f(\scalef,x) \sqrt{|\metricss|}\, dx^1dx^2dx^3\, ,
\end{equation}
with $|\metricss|=\det(\metricss_{ij})$. Physical estimates for the scale $L_0$ ranges from some tens or hundreds of Mpc \cite{MNRAS_2023_Bernui, ObservationalCosmologicalPrinciple}.

\subsection{Existence, uniqueness, and differentiability of the solutions}
\label{Sec:SolutionsProperties}

The existence and uniqueness for the smooth solution $U(\scalef,x)$ on\footnote{This is a shorthand notation for smooth on $(\scalef_m,1)\times \spacesection$ and continuous on $(\scalef_m,1]\times \spacesection$.} $(\scalef_m,1]\times \spacesection$ satisfying the initial condition \eqref{Eq:InitialConditionU} and the periodic boundary condition has been proved in Theorem 1 of Ref. \cite{LGGomes_2025_PhySc_1}, where $\scalef_m$ defines the minimal value of the scale factor above which the solution is well defined. According to its demonstration, the solution is valid for $\scalef=\scalef_1$  as far as we can ensure that $U(\scalef,x) >0$ for any $x \in \cell$ and  $\scalef_1 \le \scalef \le 1$. On the other hand, Theorem 2 therein with $s$ and $\beta$ as in \eqref{Eq:U} turns out to be the inequality
\begin{equation}\label{Eq:EnergyDensityInequalityU}
\frac{\energy_{0,m}}{3} \le \frac{1}{U^2} + K\scalef \le \frac{\energy_{0,M}}{3}     
\end{equation}
with $\energy_{0,m}$ the minimum and $\energy_{0,M}$ the maximum of initial energy distribution $\energy_0(x)$ in the cosmological cell $\cell$. By the condition \eqref{Eq:InitialConditionInequality} and the compacteness of $\cell$, there are positive constants $C_m$ and $C_M$ such that
\begin{equation}
C_m< \frac{\energy_{0,m}}{3}  - K\scalef 
\le \frac{3}{U^2} \le 
\frac{\energy_{0,M}}{3}  - K\scalef < C_M
\end{equation}
for $0<\scalef<1$. Hence, we conclude that  $\scalef_m < 0$, otherwise $(a_m,1]$ with $a_m\ge 0$ wouldn't be the maximal interval of definition for the solution. This means that $U$ is positive and smooth in an open set containing $[0,1)\times \spacesection$, even though the physical parameters as $\energy$ and $\hubble$ are not defined there, for they blow up as $\scalef \to 0$. In particular, $\ln U$ is smooth in $(\scalef_m,1]\times \spacesection$, which implies that, for any integer $n_0 \ge 0$,  the gravitational potential \eqref{Eq:GravitationalPotentialU} can be written as
\begin{equation}
\label{Eq:GravitationalPotentialSeries}
\potential  =  \ln\left( \scalef^{1/2}\, \dot{\scalef} \right) \, +\,  
\sum_{k=0}^{n_0}\, \frac{\scalef^k}{k!}\, \potential^{(k)}(x)  
+ \, \scalef^{n_0+1}\, \Phi^{({n_0})}(\scalef,x) \, ,
\end{equation}
with $\potential^{(0)}(x)=\ln U(0,x)$, 
\begin{equation}\label{Eq:GravitationalPotentialSeries2}
\potential^{(k)}(x) = \left( \frac{\partial^k\, \ln U}{\partial \scalef^k}   \right)_{\scalef=0} \, ,
\end{equation}
for $k=1,\ldots, n_0$, all smooth functions on $\spacesection$, while $\Phi^{({n_0})}$ is smooth on $(\scalef_m,1]\times \spacesection$. 

The early-time homogeneity demands that the metric \eqref{Eq:MetricConstantCurvature} comes arbitrarily close to its FLRW dust version as the scale factor approaches zero. Hence, the $0th$-order term must be constant,
\begin{equation}
\label{Eq:0thTerm}
\potential^{(0)}(x) = \textrm{const}\, ,
\end{equation}
and all the proper times of the $\normal$-observers coincide for $\scalef << 1$, 
\begin{equation}
dt \approx d\tau = \left(e^{\potential^{(0)}(x)+ \scalef\, \Phi^{(1)}(\scalef,x)}\right)\,\sqrt{\scalef}\, d\scalef
\approx C\, \sqrt{\scalef}\, d\scalef.
\end{equation}
In other words, $t \approx \tau$ and $\scalef \sim  t^{2/3}$. This is exactly the behaviour of the FLRW dust model as we approach the initial singularity, for in this regimen, the effect of the curvature is negligible and it behaves as if space were flat.

\subsection{Particle's number and  density contrast}
\label{Sec:ParticleDensityContrast}

The periodic boundary condition together with the conservation of particle number demands the total number $N$ of particles inside a cosmological cell to be conserved, that is, $\langle \pnumber \rangle = N/L_0^3\scalef^3$. Hence, denoting the particle density contrast by $\delta_p$, we have
\begin{equation}
\pnumber = \frac{N}{L_0^3\, \scalef^3}\left(\,  1 + \delta_p\, \right) 
\quad , \quad 
\langle \delta_p \rangle = 0 \, .
\end{equation}
The condition of a homogeneous early universe is written as 
\begin{equation}\label{Eq:DensityContrastAsymptoticCondition}
\delta_p \to 0 \quad \textrm{for} \quad \scalef \to 0 \, . 
\end{equation}
In this limit, the internal energy is small compared to $m_0$, so that $\pnumber$ can be readily integrated from the equation \eqref{Eq:ConservationParticleNumberU} as we put $\energy \approx m_0\, \pnumber$. Hence, for $|\energyinner| << m_0$, 
\begin{equation}\label{Eq:ParticleDensityContrast}
\delta_p = \frac{2\, L_0^3}{m_0\, N}\, \int_{0}^{\scalef}\, \gradD^2\potential(u,x)\, du \, .
\end{equation}
%
%
%
Applying the expansion \eqref{Eq:GravitationalPotentialSeries}, we get for small $\scalef$
\begin{equation}
\label{Eq:DensityContrastSeries}
\delta_p  = \frac{2\, L_0^3}{m_0\, N}\,\left( \frac{\scalef^2}{2!}\, \gradD^2\potential^{(1)}
+ \frac{\scalef^3}{3!}\, \gradD^2\potential^{(2)}+ \ldots \right)\, .
\end{equation}
%

\subsection{The total energy density}
\label{Sec:EnergyDensity}

The inequality \eqref{Eq:EnergyDensityInequalityU} written in terms of the energy density is, according to \eqref{Eq:U} and \eqref{Eq:Friedmann},  
\begin{equation}\label{Eq:EnergyDensityInequality}
\frac{\energy_{0,m}}{\scalef^{3}} \le \energy(\scalef,x) \le \frac{\energy_{0,M}}{\scalef^{3}}\, ,
\end{equation}
for $0<\scalef\le 1$. This means that $\energy \sim \scalef^{-3}$, just as in the FLRW spacetimes. The condition of early homogeneity demands it to approximate arbitrarily to its FLRW counterpart as $\scalef \to 0$, that is, 
\begin{equation}\label{Eq:EnergyDensityAsymptoticCondition}
\scalef^3\, \energy(\scalef,x) \to \mu_0 = \frac{m_0 N}{L_0^3}
\quad \textrm{as} \quad 
\scalef \to 0 \, .
\end{equation}
We apply it to set the initial condition for solving the conservation equation $\normal_\nu \grad_\mu T^{\mu\nu}=0$, 
\begin{equation}\label{Eq:EnergyConservation}
\frac{\partial\energy}{\partial \scalef} + \frac{3}{\scalef}\, \energy  
= \frac{(2+\zeta_B)}{\scalef^3}\, \left( \gradD^2\potential + |\gradD \potential|^2\right) \, .
\end{equation}
This gives us the expression
\begin{equation}\label{Eq:EnergyDensityFinalExpression}
\energy = \frac{\mu_0}{\scalef^3}  
+ (2+\zeta_B)\,\Phi(\scalef,x) \, ,
\end{equation}
with
\begin{equation}\label{Eq:EnergyDensityFinalExpression2}
\Phi(\scalef,x) = \frac{1}{\scalef^3}\,\int_{0}^{\scalef} \left(\gradD^2\potential(u,x) + |\gradD\potential(u,x)|^2\right)\, du \, .
\end{equation}
Using the expansion \eqref{Eq:GravitationalPotentialSeries} with the condition \eqref{Eq:0thTerm},  we get
\begin{equation}\label{Eq:EnergyDensityFinalSeries}
\Phi = \frac{\gradD^2\potential^{(1)}}{2\, \scalef} 
+ \frac{1}{6}\left(\gradD^2\potential^{(2)}+ 2\,  |\gradD \potential^{(1)}|^2 \right)  
+    \frac{\scalef}{24}\left(\gradD^2\potential^{(3)}+ 6\,  \gradD \potential^{(1)} \cdot \gradD \potential^{(2)} \right)  
+ \ldots \, ,
\end{equation}
where $|\gradD \Psi|^2$ and $\gradD \Psi_1 \cdot \gradD \Psi_2$ are expressions in the time-independent geometry of $\metricss$. In terms of the inner energy density $\energyinner$, we obtain for small $\scalef$
\begin{equation}
\energy = \frac{\mu_0}{\scalef^3}\left(\,  1 + \delta_p\, \right) +  
 \underbrace{\frac{\energyinner^{(-1)}}{\scalef} 
+ \energyinner^{(0)}
+ \energyinner^{(1)}\, \scalef 
+ \ldots}_{\energyinner} \, ,
\end{equation}
the coefficients being
\begin{align}\label{Eq:EnergyDensitySeriesCoefficients}
\energyinner^{(-1)} &= 
\frac{\zeta_B}{2}\, \gradD^2\potential^{(1)} \, , \nonumber \\[2mm]    
\energyinner^{(0)} &= 
\frac{\zeta_B}{6}\gradD^2\potential^{(2)} + \frac{(2+\zeta_B)}{3} |\gradD \potential^{(1)}|^2 \, , \nonumber \\[2mm]
\energyinner^{(1)} &= 
\frac{\zeta_B}{24}\gradD^2\potential^{(3)} + \frac{(2+\zeta_B)}{4}\left(\gradD \potential^{(1)} \cdot \gradD \potential^{(2)}\right) \, ,
\end{align}
and so on. In particular, the contribution of the gravitational potential to the total energy density turns irrelevant in the early universe, since $\energyinner \sim 1/\scalef$ while $\energy \sim 1/\scalef^3$, that is,
\begin{equation}\label{Eq:EnergyDensityGravitationalAsymptoticCondition}
\frac{\energyinner}{\energy} \sim \scalef^2 \to 0
\quad \textrm{as} \quad 
\scalef \to 0 \, ,
\end{equation}
which shows that our model is highly homogeneous near the initial singularity. In fact, these relations check the consistency and quantify our assumption $|\energyinner| << m_0$ used to derive formula \eqref{Eq:ParticleDensityContrast} for the early universe. This is a general tendency of the spacetimes with homogeneous and isotropic expansion, so much so that, even when we add the remaining components of the standard model, as radiation, dark energy, and cold dark matter, this behaviour still remains \cite{LGGomes_2025_PhySc_1}. In particular, it does not pose any novelty beyond the standard picture concerning the early universe, and therefore is compatible with the most accepted theories for that age, as the many facets of inflation \cite{Book_2008_Weinberg, JCAP_JMartin_2013}, for instance.

\subsection{The acceleration of expansion from the $\normal$-observers viewpoint}
\label{Sec:DecelerationCalculation}

The aceleration of expansion as it is seen for the $\normal$-observer at the fixed point $x_0 \in \spacesection$ is given by 
\begin{align*}
\frac{d^2\scalef}{d\tau^2} &= e^{-\potential(t,x_0)}\frac{d}{d t}\left(e^{-\potential(t,x_0)}\frac{d\scalef}{d t}\right)
\\
&= e^{-\potential}\frac{\partial}{\partial t}\left(e^{-\potential}\frac{d\scalef}{d t}\right)
\\
&= \scalef\, \hubble\, \frac{\partial}{\partial \scalef}\left(\scalef\, \hubble\right)
\\
&= \frac{1}{U\sqrt{\scalef}}\frac{\partial}{\partial \scalef}\left(\frac{1}{U\sqrt{a}}\right)
\\
&= -\, \frac{1}{2\,U^2\, \scalef^2} +\left(\frac{2+\zeta_B}{6\, U^3\, \scalef} \right) \frac{\partial U}{\partial s} 
\\
&= -\,\frac{1}{2\scalef^2}\left( \frac{1}{U^2}+ K\, \scalef \right) + \left(\frac{2+\zeta_B}{6\, \scalef} \right) \frac{\gradD^2 U}{U} \, .
\end{align*}
Returning with the physical variables, we have
\begin{equation}\label{Eq:Aceleration}
\frac{d^2\scalef}{d\tau^2} =
-\,\frac{\scalef\, \energy}{6} + 
\left(
\frac{2+\zeta_B}{6\, \scalef} \right)\left(\gradD^2\potential + |\gradD\potential|^2 
\right)  \, .
\end{equation}
Note that it is negative whenever the spatial variations of the potential $\potential$ are negligible, in which case it gives the well-known equation $\ddot{\scalef}=-\scalef\, \energy /6$ for the deceleration of expansion in the FLRW dust model. We postpone the discussion of the consequences of this formula to section \ref{Sec:Deceleration}.

\section{The large-scale FLRW model and the apparent dark-energy component}
\label{Sec:EffectiveLargeScaleModel}

\subsection{The large-scale FLRW model}
\label{Sec:LargeScaleFRLW}

We now turn to the fitting problem, that is, decide which FLRW model will represent the large-scale dynamics. Since our background is closely related to it, this is accomplished by two suitable gauge choices, namely, deciding which discrete subgroup of symmetries $\Gamma$ of $\metricss$ \cite{Wolf} fits better to represent the cosmological cell $\cell = \spacesection/\Gamma$ and the homogeneous scale $L_0$, and picking the gauge for the potential $\potential$, and therefore fixing the global time $t$ of the universe. We shall leave the former to be decided by observations, while for the fitting of the cosmological time, we choose the FLRW gauge, defined as  
\begin{equation} \label{Eq:FLRWGauge}
\langle e^{-2\potential}\rangle = 1 \, ,
\end{equation}
that is, the cosmic time $t$ is chosen such that
\begin{equation}\label{Eq:FLRWGauge2}
\left(\frac{\dot{\scalef}}{\scalef}\right)^2 = \langle \hubble^{2}\rangle \, .
\end{equation}
Here, we have recovered the ``dot" notation for the $t$-time derivative, $\dot{f}=df/dt$ and set $t_0$ for our current era, that is, $\scalef(0)=0$ and $\scalef(t_0)=1$. 

The large-scale expression for the total energy is given by its mean value $\langle \energy \rangle$. We adopt the critical value   
\begin{equation}\label{Eq:EnergyDensityCritical}
\energy_c = 3  \, \langle \hubble_0^2 \rangle = 3\, \left(\frac{\dot{\scalef}}{\scalef}\right)^2_{t=t_0}
\end{equation}
as our reference scale, where $\hubble_0(x)=\hubble(t_0,x)$. We introduce the cosmological parameter for the matter content
\begin{equation}\label{Eq:Omega}
\Omega_m = \frac{N\, m_0}{\energy_c\, L_0^3} \, ,
\end{equation}
just as in the standard model of Cosmology \cite{Book_2008_Weinberg}.

In the FLRW gauge, the large-scale dynamics emulates an FLRW universe with energy density given by the averaged value of \eqref{Eq:EnergyDensity}, that is,
\begin{equation}\label{Eq:EnergyFLRW}
\energy_{L}(t)= \langle \energy \rangle = \left(\frac{\Omega_m}{\scalef(t)^3} + \Omega_L(t)\right)\, 
\energy_c \, ,
\end{equation}
where, from \eqref{Eq:EnergyDensityFinalExpression} and \eqref{Eq:EnergyDensityFinalExpression2}, 
\begin{equation}\label{Eq:EnergyFLRWInner}
\Omega_L(t) = \frac{\langle \energyinner \rangle (t)}{\energy_c} = \frac{(2+\zeta_B)}{\energy_c\, \scalef(t)^3}\,\int_{0}^{\scalef(t)} \langle|\gradD\potential|^2\rangle (\scalef)\, d\scalef  
\end{equation}
appears as the averaged contribution of the local gravitational potential to the large-scale dynamics, and the effective pressure 
\begin{equation}\label{Eq:PressureFLRW}
\pressure_{L}(t) = -\, \frac{(2+\zeta_B)}{3\, \scalef(t)^2}\, \langle |\gradD\potential|^2\rangle (t)  \, ,
\end{equation}
which is negative except for the FLRW case, where it vanishes. Note that $\langle \pressure \rangle \ne \pressure_L$. In fact, from \eqref{Eq:EMTPressure},
\begin{equation}
\langle \pressure \rangle (t) = \frac{\zeta_B}{2+\zeta_B}\, \pressure_{L}(t)  \, .
\end{equation}
Note that divergent terms do not contribute to the mean values, an implication of the periodic boundary condition and Gauss integral theorem.

As we take the mean value along the equations \eqref{Eq:Friedmann}, \eqref{Eq:EnergyConservation}, and  \eqref{Eq:Raychaudhuri}, we obtain, respectively, the Friedmann equation, 
\begin{equation}\label{Eq:FriedmannAvareged}
3\, \left(\frac{\dot{\scalef}}{\scalef}\right)^2 = \energy_L  - \frac{3\, K}{\scalef^2} \, ,
\end{equation}
the equation for the conservation of energy, 
\begin{equation}\label{Eq:EnergyConservationAvareged}
\dot{\energy}_L + 3\, \frac{\dot{\scalef}}{\scalef}\, \left( \energy_L + \pressure_L\right) =0 \, ,
\end{equation}
and Raychaudhuri's,  
\begin{equation}\label{Eq:RaychaudhuriAveraged}
\frac{\ddot{\scalef}}{\scalef} =
- \, \frac{1}{6} \left( \energy_L + 3\, \pressure_L \right) \, .    
\end{equation}
It is worth noting that the FLRW equations are exactly derived from their inhomogeneous counterparts, a non-trivial fact showing that our choices are in great agreement with our intuitive notion of the cosmological principle. Therefore, on large scales, $L>>L_0$, our model is represented by the FLRW spacetime with the time function satisfying \eqref{Eq:FLRWGauge} and the same scale factor of the metric \eqref{Eq:MetricConstantCurvature}. 

In this context, it is natural to introduce the curvature contribution by $\Omega_K = -\, 3 K/\energy_c$ and the current contribution of the internal energy density by $\Omega_{L,0}=\Omega_L(t_0)$. From the Friedmann equation \eqref{Eq:FriedmannAvareged}, we get the algebraic relation  
\begin{equation}
\Omega_m + \Omega_{L,0} + \Omega_K =1 
\end{equation}
just as in the standard model, with $\Omega_m>0$, $\Omega_{L,0}\ge 0$ and $\Omega_K<1$. Moreover, the nomenclature ``critical density" in \eqref{Eq:EnergyDensityCritical} is also justified in this basis, for
\begin{equation}
 \energy_L(t_0)  = \energy_c
 \iff  
 \Omega_m + \Omega_{L,0}  =1 
 \iff  
 K=0 \, .
 \end{equation}

\subsection{The effective equation of state on large scales}
\label{Sec:LargeScaleEquationOfState}

From the large-scale point of view, the equation of state between $\energy_L$ and $\pressure_L$ is substituted by the phenomenological ansatz for the energy density given in the form 
\begin{equation}\label{Eq:EnergyFLRWSOmegaExpression}
\frac{\energy_{L}}{\energy_c} = \frac{\Omega_m}{\scalef^3} + \Omega_L(\scalef) \, ,
\end{equation}
where the function $\Omega_L(\scalef)$ is smooth in an open interval containing $[0,1]$. Pressure at large scales is, according to the FLRW conservation \eqref{Eq:EnergyConservationAvareged}, 
\begin{equation}
\frac{\pressure_{L}}{\energy_c} = -\, \frac{1}{3\, \scalef^2}\, \frac{d}{d\scalef}\left( \scalef^3\, \Omega_L(\scalef)\, \right) \, .
\end{equation}
Thus, the effective equation of state is governed by the parameter 
\begin{equation}\label{Eq:EquationOfStateEffectiveDefinition}
\eqstate (\scalef) = \frac{\pressure_L}{\energy_L} = -\, \frac{\scalef}{3\, (\Omega_m+\scalef^3\, \Omega_L(\scalef))}\, \frac{d}{d\scalef} \left(\scalef^3\, \Omega_L(\scalef)\right) \, ,
\end{equation}
that is, using the definition \eqref{Eq:EnergyFLRWInner}, 
\begin{equation}\label{Eq:EquationOfStateEffective}
\eqstate (\scalef) =  -\, \frac{\scalef\, (2+\zeta_B)\, \langle|\gradD\potential|^2\rangle (\scalef)}{3\,\energy_c\,  (\Omega_m+\scalef^3\, \Omega_L(\scalef))} \, .
\end{equation}

In the early epochs, when $\scalef$ is small, we expand all expressions around $\scalef=0$, so that we obtain from \eqref{Eq:EnergyDensityFinalSeries},  
\begin{equation}\label{Eq:DarkEnergyOmegaSeries}
\Omega_{L}(a) \approx \Omega_L^{(0)} +\Omega_L^{(1)}\, \scalef  \, .
\end{equation}
with each $\Omega_L^{(k)}$ determined by the averaged value of the corresponding internal energy coefficients in the expansion \eqref{Eq:EnergyDensitySeriesCoefficients}, 
\begin{equation}\label{Eq:Omegam0}
\Omega_L^{(0)} = 
\frac{(2+\zeta_B)}{3\, \energy_c} \langle |\gradD \potential^{(1)}|^2\rangle  
\quad \textrm{and} \quad 
\Omega_L^{(1)} = 
\frac{(2+\zeta_B)}{4\, \energy_c} \langle \gradD \potential^{(1)} \cdot \gradD \potential^{(2)}\rangle \, , 
\end{equation}
Each term is determined solely by the gravitational potential $\potential$, and vanishes in the FLRW case. Therefore, as we approach the initial singularity with $\scalef \to 0$, matter behaves according to 
\begin{equation}
\frac{\energy_{L}}{\energy_c} \approx \frac{\Omega_m}{\scalef^3} + \Omega_L^{(0)} 
\quad \textrm{and} \quad
\frac{\pressure_{L}}{\energy_c} \approx -\,  \Omega_L^{(0)}
\, ,
\end{equation}
with the equation-of-state parameter 
\begin{equation}\label{Eq:EquationOfStateEffectiveAsymtoticValue}
\eqstate (\scalef) \approx   -\, \frac{\Omega_L^{(0)}}{\Omega_m}\, \scalef^3 \, ,
\end{equation}
This is the same behaviour as the dust FLRW model with the positive cosmological constant $\Lambda = \Omega_L^{(0)}\, \energy_c$.

On the other extremity, in the late-time universe, the equation of state parameter at $t=t_0$ is
\begin{equation}\label{Eq:EquationOfStateEffectiveToday}
\eqstate_0 =  -\, \frac{(2+\zeta_B)\, \langle|\gradD\potential|^2\rangle (1)}{3\,\energy_c\,  (1-\Omega_K)}\, .
\end{equation}
Since $\langle|\gradD\potential|^2(1)\rangle = \langle|\gradD U_0|^2/U_0^2\rangle$, where $U_0(x)$ is the arbitrary initial condition \eqref{Eq:InitialConditionU}, we conclude that $\eqstate_0$ can assume any non-positive value! 
\begin{table}[htbp]
\centering
\caption{FLRW analogues}
\label{Tab:FLRWAnalogues}
\begin{tabular}{l|c}
\hline 
Epoch & FLRW analogue with  \\
& the leading term of $\eqstate(\scalef)$ \\[0.5cm]
\hline 
Early times & $\pressure = 0$ with $\Lambda = \Omega_L^{(0)}\, \energy_c$ \\[0.5cm]
\hline 
Late times  & $\pressure = \eqstate_0\, \energy$ \\[0.2cm]
&    with $\eqstate_0\le 0$ arbitrary.  \\[0.5cm]
\hline
\end{tabular}
\end{table}

We can go further and determine the higher order terms in the late-time expansion of $\eqstate(\scalef)$.
While the initial condition \eqref{Eq:InitialConditionParticle} of the distribution of matter $\pnumber_0(x)$ is constrained by the observations, the initial condition for the internal energy density, 
\begin{equation}
\energyinner_0(x)=\energy_0(x)-\pnumber_0(x) \, ,   
\end{equation}
is constrained by the determination of $\eqstate(a)$ and its derivatives at $\scalef=1$. In fact, using that $|\gradD\potential|^2(\scalef) = \gradD^2U(s(\scalef))/U(s(\scalef))$, the expressions \eqref{Eq:EnergyFLRWInner} and  \eqref{Eq:EquationOfStateEffective}, and the relation \eqref{Eq:RaychaudhuriU} for the time derivative $\partial_sU$, we can, in principle, obtain any fixed values for the first $k_0$ derivatives $\eqstate'(1), \ldots , \eqstate^{(k_0)}(1)$ by properly setting the initial condition $U_0(x)$, that is, $\energyinner_0(x)$.

A word of caution here: no mathematical proof has been presented to ensure that any range of values $\eqstate'(1), \ldots , \eqstate^{(k_0)}(1)$ can be achieved by setting the initial condition $U_0$ properly. However, it is quite general, at least. For example, let us consider the first case $k_0=1$. Define 
\begin{equation}
\omega(\scalef) = \frac{d \langle|\gradD\potential|^2\rangle}{d \scalef} 
=-\,  \frac{(2+\zeta_B)}{6} \, \left\langle \frac{\partial}{\partial s} \left(\frac{\gradD^2U}{U}\right) \right\rangle \, .
\end{equation}
Using formula \eqref{Eq:RaychaudhuriU}, we obtain 
\begin{equation}\label{Eq:EquationOfStateEffectiveToday2}
\frac{6\, \omega(1)}{2+\zeta_B} =  
\left\langle \gradD^2U_0\, \left( \gradD^2U_0+\beta U_0\right)  \right\rangle 
-  \left\langle \frac{1}{U_0} \gradD^2 \left( \left[\gradD^2U_0+\beta U_0\right]\, U_0^2 \right)  \right\rangle
\, .
\end{equation}
Denoting $\eqstate_1=\eqstate'(1)$, we get after performing the derivative of \eqref{Eq:EquationOfStateEffective} at $\scalef=1$, 
\begin{equation}
\omega(1) = \frac{3\, \energy_c\, (1-\Omega_K)}{2+\zeta_B}\, \left( \eqstate_0 + 3\, \eqstate_0^2 - \eqstate_1 \right) \, .
\end{equation}
Hence, fixing $\eqstate_0\le 0$ and $\eqstate_1 \in \R$, we can set $U_0(x)$ such that, in principle, \eqref{Eq:EquationOfStateEffectiveToday} and \eqref{Eq:EquationOfStateEffectiveToday2} attain the desired values. In other words, we can arbitrarily set the late-time relation
\begin{equation}\label{Eq:EquationOfStateEffectiveCPL}
\eqstate (\scalef) \approx \eqstate_0 + \eqstate_1\, (\scalef - 1) \, .
\end{equation}
%

\section{The conceptual and physical aspects of the tidal-dust cosmological models}
\label{Sec:PhysicalDiscussion}

\subsection{General prospects and contextualization with other approaches to cosmology}
\label{Sec:Contextualization}

The general stage of relativistic cosmology is set by choosing a suitable frame of reference with a physical meaning, which is represented by a unit timelike vector field $\normal$ pointing to the future. For our purposes here, let us admit that it is vorticity-free, that is, we split space from time such that the metric turns into
\begin{equation}\label{Eq:MetricGeneral}
\metric = -\, \underbrace{e^{2 \potential(t,x)} dt^2}_{\textrm{Proper-time}} + \underbrace{\scalef(t,x)^2}_{\textrm{Expansion}}\, \underbrace{\metrics_{k\ell}(t,x)\, dx^k\, dx^\ell}_{\textrm{Anisotropy}} \, , 
\end{equation}
with $\partial_t\det(\metrics_{k\ell}) =0$. The $\normal$-potential $\potential$ is defined up to an additive time function \cite{LGGomes_2024_CQG_1}, while the scale factor $\scalef$ is unique up to a multiplicative spatial function. When $\potential=0$, $\normal$ represents a set of free-falling observers, while $\scalef=\scalef(t)$ means that expansion is homogeneous \cite{LGGomes_2024_CQG_2}, and $\metrics_{k\ell}=\metrics_{k\ell}(x)$ means that it is isotropic. The metric \eqref{Eq:MetricConstantCurvature} represents the most general form where both expansion and spatial geometry are homogeneous and isotropic, that is, $\scalef=\scalef(t)$ and $\metrics_{k\ell}=\metricss_{k\ell}(x)$ has constant curvature. Roughly speaking, although the deep understanding of the three parts distinguished in the metric \eqref{Eq:MetricGeneral} is of fundamental importance in cosmology, here we have singled out only the first one to be analysed, while the other two were kept as simple as possible.  

The standard model of cosmology has all the special properties of the geometry of  \eqref{Eq:MetricConstantCurvature} with the extra condition of free-fall for the $\normal$-observers, $\potential=0$. The reader used to it might tend to interpret $\phi$ as a Bardeen variable \cite{bardeen1980gauge} in a very simple perturbation of a FLRW spacetime \cite{Book_2008_Weinberg}. That is not the case here, for we deal with no perturbation in this manuscript, but instead, the full set of Einstein's non-linear equations. Moreover, even if one thinks about a perturbation of "infinite" order in this gauge, that is not the case either. Conceptually, there is neither a fixed FLRW spacetime underlying ours nor any choice of "FLRW-gauge" in our approach. The metric \eqref{Eq:MetricConstantCurvature} arises as the most general possibility allowing the expansion and the spatial geometry to be both homogeneous and isotropic \cite{LGGomes_2024_CQG_2}. Note that in the early-universe epoch, $\scalef << 1$, we are indeed in the regime of small perturbations of the dust-FLRW spacetime, but as soon as the inhomogeneities grow non-linearly, our spacetime becomes highly inhomogeneous, and the subjacent FLRW structure becomes the effective large-scale model described in section \ref{Sec:EffectiveLargeScaleModel}.  

Assume that $x_v$ and $x_w$ are points in the same spatial section $t=t_0$ where $\potential$ attains a local maximum and minimum, respectively, which always exist by the compactness of the cosmological cells. For the sake of simplicity, let us also demand that they keep these properties along any instant of time, so that $\gradD\potential(t,x_w)=\gradD\potential(t,x_v)=0$ for any time $t$, that is, both are free-falling $\normal$-observers. According to the metric \eqref{Eq:MetricConstantCurvature}, during the same cosmological time interval $dt$, their proper times satisfy 
\begin{equation}\label{Eq:ProperTimeWallVsVoid}
d\tau_v = e^{\varphi(t)}d\tau_w \, ,\quad \textrm{with} \quad \varphi(t) = \potential(t,x_v) - \potential(t,x_w) >0\, .
\end{equation} 
That is, $x_v$ is in the centre of a void, where there is less matter around, and the clocks tick faster, while $x_w$ represents the middle of a clumpy region of the universe ("w" for wall), where the assembly of a large quantity of mass makes time elapse more slowly. In this context, we have three different time concepts for the description of the cosmic dynamics: $\tau_w$, $\tau_v$, and $t$. This is the starting point for the timespace cosmology \cite{Wiltshire2007NJP, Wiltshire2007PRL}. In this sense, both approaches rely on the same tenet, namely, that the differences in the proper times of the many observers and the choice of the non-physical cosmological time $t$ affect our perception of cosmic expansion. However, while the timescape approach focuses only on the large-scale phenomenological picture and leaves the local geometry to be defined by an effective "averaged" metric, ours display the exact local and large-scale metrics. In this sense, the effective FLRW model in section \ref{Sec:EffectiveLargeScaleModel} could be seen as a consistent representation of a special type of the timescape method, but differing from the usual approach in some essential points: (i) The universe expands with the same scale factor for both the void and wall regions, instead of two different ones, $\scalef_w(t)$ and $\scalef_v(t)$; (ii) The effects of spacial curvature are trivial, just as in the FLRW model, instead of a more complex picture where it vanishes near the wall and is negative in the voids; (iii) The choice for the global time $t$ was set accordingly to the gauge \eqref{Eq:FLRWGauge}, which defines $\dot{\scalef}/\scalef$ as $\sqrt{\langle \hubble^2 \rangle}$, instead of a linear combination of $\hubble(t,x_w)$ and $\hubble(t,x_v)$; (iv) The difference \eqref{Eq:ProperTimeWallVsVoid} in the proper times of the wall and void are given explicitly by a Newtonian-like gravitational potential, which plays a fundamental role in the model by connecting to their local gravitational energies, instead of a phenomenological picture relating the two times.

The spatial curvature of \eqref{Eq:MetricGeneral} will generally scale differently from its corresponding counterpart in the metric \eqref{Eq:MetricConstantCurvature}, where we have kept the same $K/\scalef^{2}$ FLRW-like behaviour. It can have a decisive role in large-scale dynamics \cite{VoidStatistics_2025_MNRAS, Wiltshire_CQG_2013, Buchert_JCAP_2013, bolejko2018emerging, Buchert_CQG_2020}, so that an observational basis has been established to test this hypothesis \cite{ClarksonBassetLuTest, EuclidTestCurvature_PRD_2014}. Within this picture, the proper-time term in \eqref{Eq:MetricGeneral} is often neglected by using a free-falling argument for the cosmological observers. However, if we decouple matter from the frame used to describe the whole universe, that is, if matter is moving in the $\normal$-frame, then $\potential$ cannot vanish, since the gravitational attraction would put in motion any physical observer initially at rest relative to it. In this case, our work establishes a relevant physical scenario for a broader discussion on whether the corresponding Newtonian-like potential could significantly contribute to the well-established effects of spatial curvature and expansion inhomogeneities. 

The principal aim of this manuscript is to exploit the conceptual consequences of the Newtonian-like gravitational forces on the large-scale dynamics, which is represented by the proper-time term in the metric \eqref{Eq:MetricGeneral}. As a necessary constraint when taking our first steps into the unknown, we have omitted many things just for the sake of simplicity, and not for their cosmological importance. From the thermodynamical point of view, for instance, we have said nothing, so that our very early universe is governed essentially by FLRW dust, which is improbable in a more realistic picture, without any mention of the CMB or the radiation and earlier eras. Even when inhomogeneities grow, and the distance to the FLRW dust model becomes way beyond the perturbative approach, we have not addressed the non-trivial consequences of the inhomogeneities in the scale factor or the many facets and consequences of anisotropy in expansion caused by the time dependence of $\metrics_{k\ell}$ \cite{LGGomes_2017_CQG, LGGomes_2021_AP,LGGomes_2022_CQG_1,LGGomes_2023_EPJC_1}.

\subsection{The Newtonian limit, orders of magnitude,  and the viscosity coefficients}
\label{Sec:InterpretationViscosityParameters}

The Newtonian potential $\potential_N$ satisfies the Poisson equation $\gradD^2\potential_N = m_0\pnumber/2$ at the time $t=t_0$ with a typical null boundary condition at "infinity". The $\normal$-potential $\potential$ is its analogue in our model and satisfies Raychaudhuri's equation \eqref{Eq:Raychaudhuri} coupled to the rest of Einstein's equations under the periodic boundary conditions of section \ref{Sec:BoundaryConditionPeriodic}. The analogy between $\potential_N$ and $\potential$ was essential to interpret $\gradD\potential$ as a local gravitational force and $\tide_{k\ell}$ as the Newton-like gravitational tidal tensor. However, Raychaudhuri's equation reduces to Poisson's only if we put our local universe in a dome and ignore the cosmological effects on it, by setting $\potential <<1$, $\pressure/\energy << 1$, and $\hubble \approx 0$. This is a misleading approach to investigate our model in a cosmological perspective, for Einstein's equations put $\energy$, $\hubble^2$, and $|\gradD\potential|^2$ on the same order of magnitude. Hence, in a relativistic cosmic scenario, we can have $\potential_N<<\potential$. Moreover, since $\potential$ also quantifies the difference in proper time between voids and walls, as in \eqref{Eq:ProperTimeWallVsVoid}, the timescape estimates for the clock-rate relation suggest a significant role for the magnitude of $\potential$ \cite{wiltshire2009average}, contrary to the Newtonian expectations coming from the supposition that $\potential \approx \potential_N$.

If we allow anisotropies in the context where spatial geometry is homogeneous and isotropic, the traceless spatial Einstein's equations (eq. (30) in Ref. \cite{LGGomes_2024_CQG_1} with $\sigma^k_i= C\, \hubble\, \Sigma^k_i$) with the condition \eqref{Eq:EMTAnisotropicStress} is
\begin{equation}
e^{-\potential}\, \partial_t\sigma^k_i + 3\, \hubble\, \sigma^k_i 
= \left( 1- \zeta_{A} \right)\, \tidea_{ij}\, ,
\end{equation}
where $\sigma^k_i$ is the shear tensor of $\normal$. Therefore, $\zeta_A \ne 1$ makes $\tidea_{ij}$ work as a source of anisotropy, so that the condition \eqref{Eq:TideCurvatureStressEquilibriumZeta} is nothing but the compatibility with our initial assumption of isotropy, that is, $\sigma^k_i=0$. In particular, in the context of \eqref{Eq:ViscosityCoeficientsMaximum}, it demands
\begin{equation}
\zeta_{A} =\zeta_{max} =1 \quad, \quad  0 \le \zeta_{B} \le 1  \quad \textrm{,} \quad \pi^{(LPF)}_{k\ell} = 0\, .  
\end{equation}
From the point of view of the $\normal$-observers, this says that the matter distribution presents no resistance against the shear induced by the gravitational tides. In particular, this gives a physical realization for the tide-curvature-stress equilibrium in Ref. \cite{LGGomes_2024_CQG_2, LGGomes_2025_PhySc_1} and the condition $\pi_{ij}=-2\, E_{ij}$ involving the electric part of the Weyl tensor \cite{LGGomes_2021_IJMPD,LGGomes_2022_CQG_2}.

\subsection{The deceleration of the expansion}
\label{Sec:Deceleration}

In the early period, the universe decelerates from the viewpoint of every $\normal$-observer, for according to \eqref{Eq:GravitationalPotentialSeries}, \eqref{Eq:0thTerm},  \eqref{Eq:EnergyDensityAsymptoticCondition}, and \eqref{Eq:Aceleration},     
\begin{equation}
\frac{d^2\scalef}{d\tau^2} \approx -\, \frac{\mu_0}{6\, \scalef^2} < 0 \quad \textrm{(Early times)}
\, .  
\end{equation}
This is the same expression as in the FLRW dust model. In late times, this scenario is maintained for the observers in the middle of the voids, at least. In fact, in the centre of a void today, where the function $\potential(t_0, \cdot)$ has a local maximum, say, at $x_v\in \spacesection$, we have $\gradD\potential(t_0,x_v)=0$ and $\gradD^2\potential(t_0,x_v) \le 0$. Hence, the positiveness of $\energy$ ensures that the expression \eqref{Eq:Aceleration} is negative, that is, 
\begin{equation}\label{Eq:AcelerationVoid}
\frac{d^2\scalef}{d\tau_v^2} <0 \qquad \textrm{in a void.}  
\end{equation}
Moreover, we should expect the voids to become more homogeneous as time passes, so that both $\energy$ and $\gradD\potential$ vanish asymptotically there. Hence, at this epoch we have $\potential(a,x) \approx \potential(x_v)$ for any observer therein. In this case, they will witness an asymptotic FLRW universe with the time function $t= e^{-\potential(x_v)}\tau_v$, $\tau_v$ its proper time, with an open universe for $K\le 0$, and closed for $K>0$. Since the scale factor is the same for any kind of $\normal$-observers, we conclude that this situation is valid for any late-time observer far enough from a clumpy region.

On the other hand, there might be some $\normal$-observers to whom the expansion is accelerated. For them, time passes slower than in the voids, for they are close to dense agglomerations of matter. Whether this kind of $\normal$-observer exists or not deserves further investigation that is beyond our scope here. However, we can work out some estimates if we think in a Newtonian way and, for the sake of simplicity, set $\zeta_B=0$. In this case, we assume the centre of an overdense region to be a local minimum of the potential $\potential(t_0, \cdot)$, say $x_w \in \spacesection$. If we assume the Newtonian Poisson equation $\gradD^2\potential(t_0,x_w) \approx \mu_ w/2$, where $\mu_w$ is the rest-mass density at $x_w$ when $\scalef(t_0)=1$, with $\gradD\potential(t,x_w) = 0$ for all $t$, and use the equations \eqref{Eq:EnergyDensityFinalExpression}, where we compute the integral defining $\Phi(t_0,x_w)$ as $\alpha\, \mu_w/2$, for some $0 < \alpha<1$, and apply it to the acceleration \eqref{Eq:Aceleration}, we estimate that the observer in an overdense region will perceive an accelerated expansion when
\begin{equation}
\mu_w \gtrsim \frac{\mu_0}{1-\alpha} \, . 
\end{equation}
This is a rough estimate telling us that there must be a critical value $0 < \alpha_c<1$ for which any $\normal$-observer at an agglomeration of mass with density above $\frac{\mu_0}{1-\alpha_c}$, $\mu_0$ the average mass density, must measure acceleration of expansion, despite the deceleration observed in the voids. However, even if such an overdense region forms at some late time, the situation is completely distinct from the accelerating FLRW scenario. In fact, this is only an apparent effect that arises as mass is being poured on that region at the cost of slowing down the clock's pace of the observers therein.

\subsection{Exactness of the fitting problem and the acceleration of the large-scale FLRW model}
\label{Sec:LargeScaleApparentAcceleration}

In section \ref{Sec:EffectiveLargeScaleModel}, we came up with a solution to the fitting problem that could be called exact: the effective large-scale model obeys the very same equations as a FLRW one! The difference between the two is that there is no freedom of choice for the equation of state involving the large-scale energy density $\energy_L$ and pressure $\pressure_L$, defined in \eqref{Eq:EnergyFLRW} and \eqref{Eq:PressureFLRW}, respectively, since the equation-of-state parameter \eqref{Eq:EquationOfStateEffectiveDefinition} is a consequence of the physical assumptions for the background model. Note also that $\energy_L > 0$, while $\pressure_L \le 0$. This last condition comes from the second term in \eqref{Eq:EMTPressure} defining the background pressure, which in turn results from the particle frame being accelerated relative to that of homogeneous and isotropic expansion, as seen in \eqref{Eq:ChangeOfFrameStressTensor}.

The relations obtained so far allow us to write explicitly the coefficients necessary to construct the luminous distance relations just as they are obtained in the standard model. Since the current deceleration parameter in the FLRW case is $q_0=(\energy_L(t_0)+3\pressure_L(t_0))/2\energy_c$ \cite{Book_2008_Weinberg, VisserCQG2004jerk}, we get
\begin{equation}\label{Eq:DecelerationParameter}
q_0 = \frac{1}{2} \left( 1- \Omega_K \right)\,  \left( 1+3\, \eqstate_0 \right)
\end{equation}
Dark energy is the main candidate to explain the observed luminous distance curve $d_L(z)$ \cite{Book_2008_Weinberg}, where $q_0$ is the value determining the coefficient accompanying $z^2$ in the Taylor expansion in the redshift $z$ around $z=0$. It is usually associated with the cosmological constant $\Lambda$, a fluid with the equation of state $\eqstate(\scalef)=-1$. Therefore, the arbitrariness of the negative $\eqstate_0$, as shown in \eqref{Eq:EquationOfStateEffectiveToday}, could explain the constant dark energy as an apparent effect arising from the local gravitational field. This would put us in an apparently contradictory situation: while the luminous distance observations indicate acceleration, the universe decelerates for the $\normal$-observers, as we have seen in section \ref{Sec:Deceleration}. This conundrum is explained by the fact that the global time $t$ coming from the FLRW gauge \eqref{Eq:FLRWGauge} cannot be realised as the proper time $\tau$ of the observers seeing a homogeneous and isotropic expansion. A similar behaviour can be found in a timescape basis \cite{wiltshire2009average, wiltshire2014cosmic}, where the "clock ansatz" plays the role of our "gauge choice" for $t$. 

The corrections up to the first order for $\eqstate_0$ are obtained by the truncation \eqref{Eq:EquationOfStateEffectiveCPL},  first proposed by Chevallier, Polarski, and Linder (CPL) \cite{IJMPD_2001_Chevallier, PRL_2003_Linder}. It plays an important role in connecting the theory with observations \cite{AP_2006_Linder, Nature_2017_CPL}. Among the many possibilities of explications for it~\cite{Amendola:2007,Guo:2007,Wang:2016, Guo:2007, Wang:2016,Pradhan:2020,PRD_2015_Scherrer, PLB_2006_CPL_classification}, we attribute the rise of $\eqstate_0 < 0$ and $\eqstate_1$ to the mean value of the inhomogeneities of the local gravitational field. Moreover, an arbitrary truncation in order higher than one in the equation of state parameter $\eqstate(a)$ can also be obtained by properly defining the initial condition $\energyinner_0(x)$, with $\eqstate_0 \le 0$ being the only apparent constraint. Hence, our approach is also suitable for extensions of the CPL formalism to higher orders \cite{Nesseris2025cpl, JCAP_2025_Gonzalez}. Since this is directly related to the different time scales in the model, which we have heuristically represented by $\tau_v$ (voids), $\tau_w$ (walls), and $t$ (cosmological) in section \ref{Sec:Contextualization}, it is not surprising that formula \eqref{Eq:EquationOfStateEffective}, their Taylor coefficients, and the follow-up interpretation are all very similar to their well-established timescape analogues \cite{wiltshire2009average, wiltshire2014cosmic}. 

It is not clear to what extent the large-scale FLRW luminous distance should be valid. Due to the subtleties underlying the role of inhomogeneities in such cosmological observations \cite{LGGomes_2021_IJMPD, CliftonFerreira_2009, bentivegna, CQG_2010_Maartens_Clarkson, JCAP_2009_LumDistance_Rasanen, JCAP_2010_LumDistance_Rasanen, koksbang2019towards, koksbang2021understanding}, further clarification is needed. This is an important issue that will be considered elsewhere.  

There is a huge body of literature accounting for the many possibilities for the nature and properties of the dark-energy sector \cite{DarkEnergy_Copeland, Book_DarkEnergy_Amendola, CRP_JMartin_2012, Wiltshire2007NJP}. In particular, the timescape scenario has been put under observational scrutiny with prominent results and expectations\cite{lane2025cosmological, seifert2025supernovae}. Our proximity with the latter and the arbitrariness of the effective equation of state \eqref{Eq:EquationOfStateEffective} are two valuable features for testing these ideas elsewhere in the future. However, fitting real data is not in the scope of this investigation, but instead, we have focused in enhancing our knowledge of the different roles the inhomogeneities can have in the description of the universe at large scales. In particular, it supports that backreaction can be big enough to account for the observed universe \cite{Buchert_2015}, at least in principle, contrary to its denial \cite{Green_2014}.

\section{Conclusion}
\label{Sec:FinalRemarks}

In this manuscript, we have introduced a cosmological model with matter inhomogeneously distributed in space as part of an effort to expand our mathematical understanding of the cosmological principle, namely, models beyond the FLRW spacetimes that are homogeneous and isotropic on large scales. We have focused on novelties for the late-time regime, while we kept the earlier epochs as homogeneous and isotropic as in the standard picture of Cosmology. The matter content has been described phenomenologically as an ensemble of particles whose number is conserved and the local centre of mass is falling under the gravitational force therein, with no resistance to the shear caused by gravitational tides. We have shown that the problem of Einstein's equations in this case is well-posed, and each solution gives rise to an FLRW spacetime on large scales, where the equation of state parameter has the form
\begin{equation}\label{Eq:EqStateGeneralTaylorSeries}
\eqstate (\scalef) = \eqstate_0 + \eqstate_1\, (\scalef-1) + \frac{\eqstate_2}{2!}\, (\scalef-1)^2 + \ldots \, ,
\end{equation}
with $\eqstate_0\le 0$ and each of the coefficients $\eqstate_i$ determined by the current distribution of the energy and particle densities along space. This turns out to be a significant result in the literature of backreaction in Cosmology for the following reasons:(i) It is an explicit model where the late-time acceleration of the universe appears as an apparent effect caused by the local gravitational field; (ii) An effective FLRW spacetime appears naturaly on large-scales; (iii) The coefficients in the Taylor series \eqref{Eq:EqStateGeneralTaylorSeries} are quite arbitrary to fit the observational data; (iv) The deceleration parameter is negative for $\eqstate_0<-1/3$, indicating a large-scale apparent acceleration, while the observers who witness an homogeneous and isotropic expansion are typically decelerating relative to each other. In particular, this last feature tells us that the global time used to describe the large-scale dynamics cannot be realised as the proper time of almost any observer to whom the expansion and geometry are homogeneous and isotropic, except in the FLRW case, of course. 

The mathematical implementation of the cosmological principle, in the sense that the universe is homogeneous and isotropic on large scales, is not exclusive to the FLRW spacetimes. Models with homogeneous and isotropic expansion \cite{LGGomes_2024_CQG_2} have a lot to contribute to theoretical Cosmology by realising physically meaningful universes where this tenet is also valid. Their conceptual development and subsequent observational scrutiny could turn them into a suitable mathematical framework underlying our explanation of the Cosmos.   
\\[3mm]

{\bf Acknowledgements:} The author is thankful for the support from FAPEMIG, project number RED-00133-21.\\[4mm]

\bibliography{ref_IHIS.bib}

\end{document}